\documentclass{mem}
\usepackage{natbib}\usepackage{txfonts}\usepackage{balance}
\usepackage{graphicx}
\usepackage[a4paper]{hyperref}
\idline{75}{1}
\begin{document}
\def\teff{$T\rm_{eff }$}
\def\kms{$\mathrm {km\ s}^{-1}$}

\title{
A close look to quasar-triggered galaxy winds
}

\subtitle{
Is the black hole--bulge relation self-regulated?
}

\author{
P. \,Monaco\inst{1,2} 
          }
  \offprints{P. Monaco}

\institute{
Dipartimento di Astornomia -- Universit\`a di Trieste, 
Via Tiepolo 11,
I-34131 Trieste, Italy
\and
Istituto Nazionale di Astrofisica --
Osservatorio Astronomico di Trieste, Via Tiepolo 11,
I-34131 Trieste, Italy
\email{monaco@ts.astro.it}
}

\authorrunning{P. Monaco}

\titlerunning{Self-regulated BH--bulge formation}

\abstract{ 
We discuss the role of feedback from AGNs on the formation of
spheroidal galaxies.  The energy released by an accreting Black Hole
(BH) may be injected into the ISM through blast waves arising directly
from the central engine, radiation pressure or radiative heating.  A
scenario is described in which radiative heating perturbs the
methabolism of a star-forming spheroid, leading to a critical stage
where SNe form a cold expanding shell, pushed out of the galaxy by
radiation pressure from the AGN.  This mechanism can regulate the
BH--bulge relation to the observed value.  However, this relation may
be simply imprinted by the mechanism responsible for the nearly
complete loss of angular momentum of the gas that accretes onto the
BH.  Using a novel model of galaxy formation that includes AGNs, we
show that models without self-regulation have problems in reproducing
the correct slope of the AGN luminosity function, while models with
winds give a much better fit; however, all these models are almost
indistingishable as far as their predicted BH--bulge relation is
concerned.  Finally, we show that the downsizing of the faint AGNs is
most likely due to kinetic feedback in star-forming bulges.
}

\maketitle{}

\section{Introduction}

AGN feedback is now considered as a fundamental process for the
formation of galaxies.  In particular, the correlation between the
mass of the super-massive BHs, remnants of the past quasar epoch, and
that of their host spheroids (ellipticals and spiral bulges) has often
been proposed to be caused by self-regulated feedback by an accreting
BH, experienced at the formation epoch of the host itself (see, e.g.,
Ciotti \& Ostriker 1997; Silk \& Rees 1998; Haehnelt, Natarajan \&
Rees 1998; Fabian 1999; Granato et al. 2001, 2004; Murray, Quataert \&
Thompson 2005; Sazonov et al. 2005; Lapi, Cavaliere \& Menci 2005;
King 2005).  However, many authors have modeled the joint AGN/galaxy
formation, reproducing the BH--bulge correlation by assuming a
relation between star formation in the host and accretion onto BHs
(see i.e. Kauffmann \& Haehnelt 2000; Cavaliere \& Vittorini 2002;
Mahmood, Devriendt \& Silk 2004; Bromley, Somerville \& Fabian 2004;
Bower et al. 2005).  In other words, the BH--bulge relation could be
due either to the self-limiting action of a wind or to the mechanism
responsible for the nearly complete loss of angular momentum of the
accreting gas. In this paper we will try to clarify this point,
following closely the recent paper by Monaco \& Fontanot (2005).

Star formation in galaxies and BH accretion show some analogies.  In
both cases a necessary but not sufficient condition to trigger these
processes is the presence of cold and dense gas within a galaxy.  Then
this gas needs either to gather into massive clouds (say, molecular
clouds) or to lose its angular momentum.  After this, star formation
or accretion can take place, and the resulting energy feeds back on
the collapsing/accreting matter, thus regulating the process.  So,
while feedback is a fundamental step in the process, the real
bottleneck of the process is either the formation of star-forming
(molecular) clouds or the loss of angular momentum.  These processes
must be properly taken into account.

Regarding AGNs, many pieces of evidence highlight the need for
feedback.  The ``anti-hierarchical'' behaviour, also called
``downsizing'', of the AGN population (Hasinger et al. 2005; see also
Merloni 2004, Marconi et al. 2004), strengthened by the observed
dearth of faint AGNs in the GOODS fields (Cristiani et al. 2004;
Fontanot et al. 2006a), points to an early assembly of the most
massive BHs and a correspondingly later assembly of smaller BHs, at
variance with the hierarchical trend of DM halos.  A similar trend is
suggested to be present in elliptical galaxies (Treu et al. 2005).
The hierarchical order can be reverted by feedback (see, e.g., Granato
et al. 2001, 2004; Bower et al. 2006); however, whether this feedback
has a stellar or AGN origin is not easy to understand.  In particular,
reverting the hierarchical order requires on the one hand to delay the
formation of small objects, and on the other hand to prevent the large
objects formed at early time to continue accreting matter for a Hubble
time.

\section{Injection mechanisms}

The AGN releases a huge amount of energy, roughly two orders of
magnitude larger than that required to unbind a typical bulge, but
most of it comes out in the form of radiation or in highly collimated
jets.  Whether this energy can be injected into the ISM so as to
trigger a massive galactic wind, able to wipe out the galaxy, is still
debated (see Begelman 2004 for a review).  A first possibility,
discussed in length by King, Cavaliere and De Zotti in this
conference, is related to injection of kinetic energy directly from
the central engine of the AGN.  Such events are observed in BAL
quasars, which are very common at high redshift; however, the effect
of such winds on the ISM depends on the poorly known covering factor
and geometry of the wind, and on the amount of mass of the ejecta.
These quantities are still uncertain, so while such winds are
perfectly plausible, it is useful at the moment to keep an open mind
and consider other possibilities.

The jets emitted by slowly accreting BHs (in units of the Eddington
accretion) are another very important mechanism, but due to the high
level of collimation their direct effect is irrelevant for the galaxy
(actually, the longly known alignment effect of star formation and
jets in radio galaxies tells us that jets stimulate star formation
more than quenching it).  On the other hand, these are the most
promising candidates to quench the cooling flows in galaxy clusters,
and, as pointed out by Frenk in this conference, this quenching is
fundamental to keep the most massive galaxies old.  So, jet feedback
from AGNs is probably very important to solve half of the downsizing
problem, that of preventing the most massive galaxies to continue to
form stars at late times.

Radiation arising from the accreting BH exherts pressure on the ISM,
and can cause runaway radiative heating of the cold phase to a
temperature of the order of the inverse-Compton temperature of the
quasars, $\sim10^7$~K (Begelman, McKee \& Shields 1983).  Radiation
pressure alone can lead to a complete removal of ISM only when it
amounts to no more than 5-10 \% of the bulge mass (Murray, Quataert \&
Thompson 2004), while radiative heating alone can wipe out the cold
phase of an elliptical when the gas amounts to a few \% of the mass
(Sazonov et al. 2005).  On this basis, it is fair to conclude that a
shining quasar does influence the galaxy only in a marginal (though
important) way.

\section{A possible interplay with SNe}

Monaco \& Fontanot (2005) proposed a mixed scenario where SNe and the
AGN cooperate to produce a massive wind able to self-regulate galaxies
and BHs.  They started by asking what is the effect of a shining AGN
on the methabolism of a star-forming spheroid.  Using the model of
Monaco (2004) as a starting point, they noticed that the
inverse-Compton temperature of a quasar is similar to the temperature
of the hot phase predicted to be present in a star-forming thick
structure.  Indeed, multiple SN explosions associated with a single
star-forming cloud give rise to a single super-bubble; in thin systems
like spiral discs the super-bubbles manage to blow out of the galaxy,
but in thicker or denser systems this is not the case, and the
efficiency of injection of (thermal and kinetic) energy into the ISM
is very high.  The resulting ISM is then highly pressurized and the
hot phase is very hot ($T\sim10^7$~K).  In this case super-bubbles are
typically pressure-confined by the hot phase before the blast can cool
and form a Pressure-Driven Snowplough (PDS; Ostriker \& McKee 1988;
Monaco 2004).

A shining quasar adds to the system an evaporative mass flow which
moves mass from the cold to the hot phase.  When this mass flow
becomes significant with respect to the star-formation rate, the hot
phase becomes denser, and the blasts can get to the PDS regime.  The
snowploughs then compress the hot gas and collapse it back to the cold
phase; the system tries to get back to the original configuration.
This leads to a drop in thermal pressure, so that the blasts become
pressure-confined at larger radii.  In typical situations this leads
to a percolation of these cold shells, and thus to the formation of a
super-super-bubble, or, in other words, to a galaxy-wide cold
super-shell.

This cold and thick shell is then pushed by radiation pressure.  Its
opacity will be initially high, due to the (possibly) large mass
involved and to the likely presence of dust, which in this conditions
is dynamically coupled to the gas (Begelman 2004).  The efficiency of
energy injection is then simply $\sim v/c$, where $v$ is the speed of
the outwardly moving shell.  Its initial velocity is estimated to be
of order $\sim200$ \kms, which is the final speed of PDS's at the
percolation time.  As shown in Monaco \& Fontanot (2005), such a shell
is accelerated by radiation, but the work done on it is roughly equal
to the gravitational energy it gains, so the velocity remains roughly
constant.  Eventually, the column density and the opacity of such a
shell will decrease up to a point that radiation pressure is be
negligible.  This will take place typically when the gas is already
out of the galaxy; however, an ejection of this gas out of its DM halo
is extremely unlikely, as the gas should snowplough all the hot halo
gas in a condition where radiation pressure is not efficient any
more. For a blast velocity of $\sim200$ \kms, the hot halo phase would
immediately pressure-confine the blast.

During the ejection of the shell, radiative heating is irrelevant due
to the high pressure of the PDS (radiative heating is relevant only
when radiation pressure is much higher than the ISM pressure).  The
drop in pressure consequent to the stalling of the blast would make
radiative heating efficient again, so that this gas would be quickly
heated up.  Such a shell would then be difficult to observe: it would
obscure both the quasar and the galaxy during the ejection (an
efficient radiation pressure implies a high opacity), then it would be
visible as a slow warm absorber, with a systemic velocity of $\sim200$
\kms, during the evaporation phase.  It would be not easy to
distinguish this component from all the complex structure of
absorption lines associated with a typical quasar, though the mass
associated with the object would be much larger.

Considering a bulge with mass $M_{\rm bul}$, normalized to $10^{11}$
M$_\odot$, harbouring a BH accreting at a rate $\dot{M}_{\rm BH}$,
normalized to 4 M$_\odot$ yr$^{-1}$ (the Eddington accretion rate of a
$1.6\times 10^8$ M$_\odot$ BH typically hosted in the $10^{11}$
M$_\odot$ bulge), the maximum amount of mass that radiation pressure
can eject, quantified as a fraction $f_s$ of the total bulge mass, is
estimated as:

\begin{equation}
f_s = 0.21 
\left(\frac{\dot{M}_{\rm BH}}{4\ {\rm M}_\odot\, {\rm yr}^{-1}}\right)^{1.5}
\left(\frac{M_B}{10^{11}\ {\rm M}_\odot}\right)^{-1.65}
\label{eq:f_s} \end{equation}

\noindent
Then, this mechanism is able to self-regulate the BH--bulge system by
hampering BH accretion when it overshoots the local BH--bulge relation.

Another important consequence of this wind model is connected to the
fact that the wind is generated throughout the galaxy (or at least in
its inner regions).  In this case a fraction of the matter is expected
to be collapsed to the centre by the same blasts that created the
outward-moving shell.  This would give rise to a wind-stimulated
episode of quasar shining.  This would then be the main visible phase
of the quasar, as it would presubably reach its peak luminosity when
the shell has already been destroyed.  A more precise assessment of
this point is very difficult because the three timescales of ejection
of the shell, loss of angular momentum of the gas and accretion
timescale are all very similar.

It is then fair to distinguish between the ``dry wind'' scenario,
where the kinetic energy coming from the AGN drives a massive removal
of ISM, and this ``accreting wind'' scenario, where the wind,
triggered throughout the galaxy, stimulates a further accretion
episode.  Whether such an indirect and complicated mechanism is really
in place is not easy to assess, and there is no reason to conclude
that such a mechanism represents a part of reality.  However, the idea
that quasar shining perturbs significantly the methabolism of the ISM
is worth pursuing, and the generation of a galaxy-wide super-shell
comes from the Monaco (2004) model very naturally, without any tuning
of parameters.

\section{Self-regulated BH--bulge relation?}

As mentioned in the Introduction, the BH--bulge relation can be
generated either by imposing a correlation between star formation (in
bulges) and loss of angular momentum, or by letting the BH--bulge
system to self-regulate.  However, within the context of a galaxy
formation model, the amount of gas available to the BH must be
specified; in most cases (see the references in the Introduction) the
accretion rate is related to the star-formation rate, so that the
BH--bulge relation is imprinted by the (unknow) mechanism responsible
for the loss of angular momentum.  In other words, there is no real
need of self-regulation to justify the BH--bulge relation.

\begin{figure*}[t!]
\resizebox{\hsize}{!}{\includegraphics[clip=true]{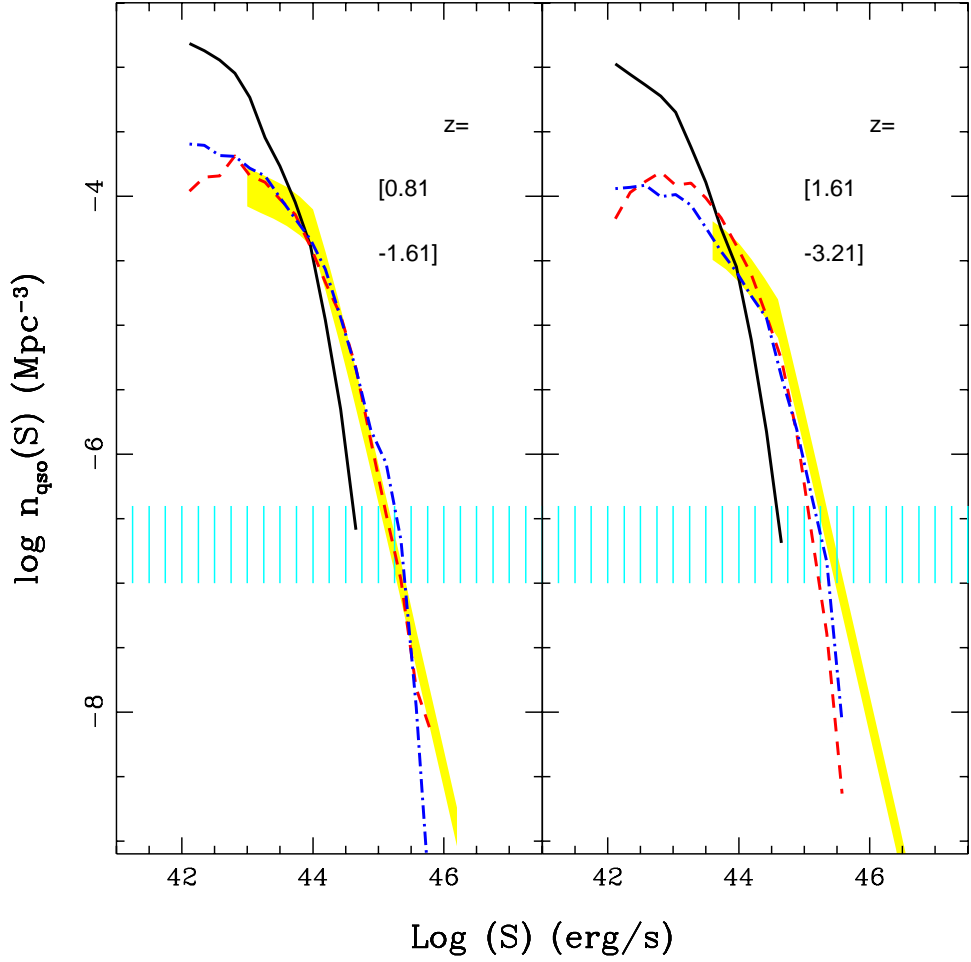}}
\caption{\footnotesize Predicted AGN LF in the hard-X band (2-10 keV),
compared with the analytic fit of Ueda et al. (2003) (yellow region).
The redshift range is indicated in the panels. Continuous black lines
correspond to the model without quasar-triggered winds, red dashed and
blue dot-dashed lines correspond respectively to the model with dry
and accreting winds.  The hashed area highlights the region where the
statistics of the model is poor.  The low-redshift LFs are not shown
because they are very sensitive to the uncertain quenching of the
cooling flow; see Fontanot et al. (2006b) for details.  }
\label{hxlf}
\end{figure*}

\begin{figure*}[t!]
\resizebox{\hsize}{!}{\includegraphics[clip=true]{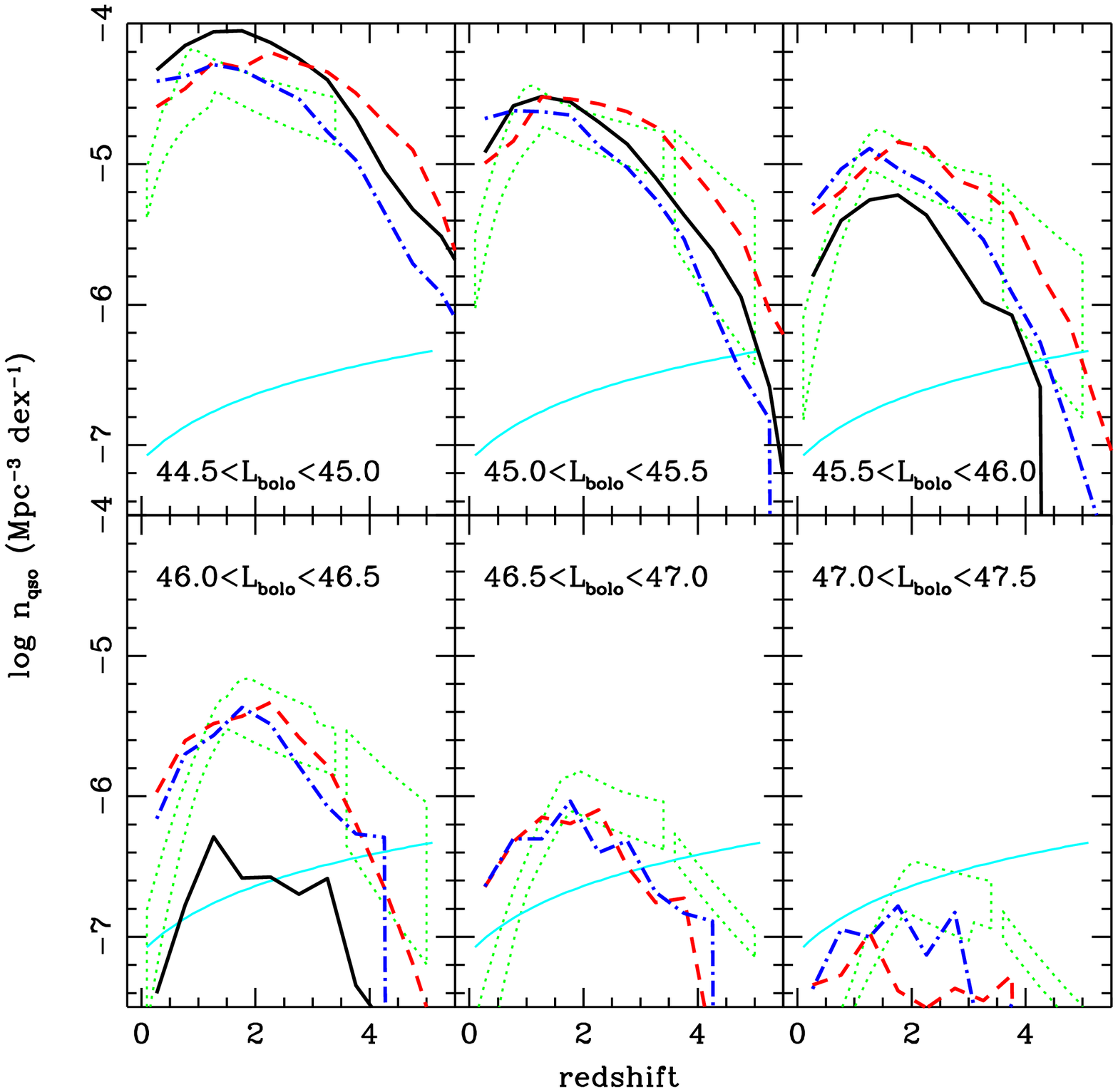}}
\caption{\footnotesize Predicted number density, as a function of
redshift, of AGNs in bins of bolometric luminosity.  The region
allowed by data (based on Cristiani et al. 2004; Ueda et al. 2005; La
Franca et al. 2005) is highlighted by green dots.  Lines refers to
models as in figure~\ref{hxlf}.  The cyan line marks the regions where
the statistics of the model is poor.  }
\label{numdens}
\end{figure*}

This of course does not imply that no self-regulation is in place.
Monaco \& Fontanot (2005) proposed to connect self-regulation to the
amount of mass available for accretion: whenever this is higher than
that required to reproduce the BH--bulge relation at $z=0$,
self-regulation can limit the BH mass to the required value. 

We have included BH accretion into a complete model for galaxy
formation, quickly described in Monaco \& Fontanot (2005); this model
will be presented very soon (Monaco, Fontanot \& Taffoni 2006).  We
show here some new results (Fontanot et al. 2006b, in preparation)
that highlight a possible role of quasar-triggered winds.

The free parameters of the model are set by requiring good fits for
the galaxy population; in particular, this model is able to reproduce
correctly the early assembly and late passive evolution of the most
massive galaxies.  With the standard set of parameters, that include
no quasar-triggered winds and then no self-regulation of the BH-bulge
systems, we find it impossible to fit the observed quasar LFs.  This
is shown in figure~\ref{hxlf}, where we compare models and data in
terms of hard X-ray LFs (Ueda et al. 2003); the black line,
representing the model with no quasar-triggered winds, is too steep.
This model is however able to reproduce the $z=0$ BH--bulge relation,
which shows that this relation is not a very strong constraint after
all.  The only way to obtain reasonable fits is by increasing the
number of degrees of freedom, i.e. by allowing for quasar-triggered
winds and self-regulation.  This is done at the cost of not having a
unique solution; in fact, we identify two possible solutions, one with
``dry winds'', the other with ``accreting winds''.  The predicted
evolution of the number density of AGNs is presented in
figure~\ref{numdens}, where it is evident that the downsizing of the
AGN population is reproduced.

\subsection{Stellar kinetic feedback causes downsizing}

\begin{figure*}[t!]
\resizebox{\hsize}{!}{\includegraphics[clip=true]{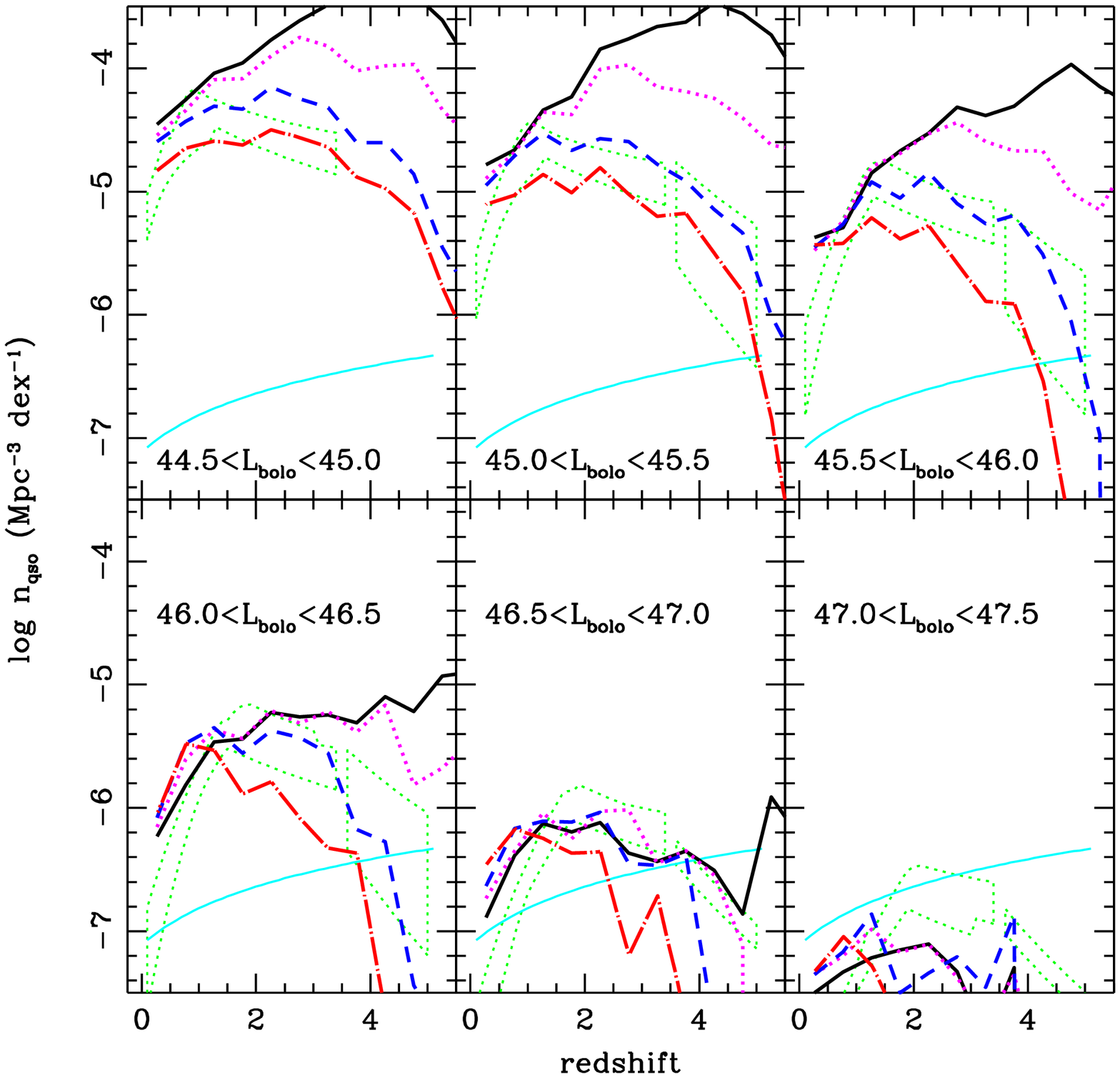}}
\caption{\footnotesize As in figure~\ref{numdens}, for models with dry
winds and $\sigma_0=0$, 40, 80 and 120 \kms.  }
\label{vturb}
\end{figure*}

We show here that the origin of the downsizing of AGNs in our model is
mostly due to kinetic feedback in star-forming bulges.  To remove gas
from a galaxy with stellar feedback there are two main ways, namely to
heat the gas (thermal feedback) or to accelerate it (kinetic
feedback).  In the model of Monaco (2004) for a star-forming ISM, gas
is heated by SNe to a temperature ranging from $10^6$ K, typical in
spiral discs, to $10^7$ K, typical in thick systems like star-forming
bulges.  This gas can easily escape the galaxy, with the exception of
the most massive bulges.  This leads to hot wind rates that are very
similar to the star formation rate.  On the other hand, the momentum
given by SNe to the ISM can accelerate clouds to high velocities.
This does not take place in discs, where the velocity dispersion of
clouds regulates to $\sim6$ \kms, but can happen in thick star-forming
systems, where the super-bubbles cannot blow out of the system and so
the kinetic energy injected into the ISM is much higher.  If the
energy injection rate from SNe is equated to the loss rate by the
decay of turbulence (whose timescale is very similar to the crossing
time), it is very easy to obtain that the velocity dispersion of
clouds $\sigma_{\rm cold}$ scales with the star-formation time-scale
$t_\star$ as:

\begin{equation}
\sigma_{\rm cold} = \sigma_0 t_{\star}^{-1/3}
\label{eq:turb}\end{equation}

\noindent
where $t_\star$ is given in Gyr and $\sigma_0$ in \kms.  This
mechanism is very effective in removing mass from small star-forming
bulges.  In figure~\ref{vturb} we show the results of the dry wind
model as a function of $\sigma_0$; kinetic feedback is able to delay
the peak of activity of faint AGNs, while bright quasars remain mostly
unaffected.

\section{Conclusions}

Star-forming galaxies at high redshift harbour shining AGNs, and the
energy emitted by them can indeed affect the evolution of the galaxy;
however, how this happens is still very unclear.  We have mentioned
many plausible ways in which some of the AGN energy could be injected
into the ISM; as typical cases, the energy could come directly from
the AGN in form of a powerful blast, or could be generated throughout
the galaxy, say by SNe, then pushed away by radiation pressure.  In
the first case star formation and accretion would be quenched, in the
second case a secondary accreetion event would be induced.  In all
cases, the efficiency of energy injection is unlikely to be higher
than 0.5 \%.

The BH--bulge relation is found to be not a very strong constraint
for the joint formation of bulges and BHs.  Indeed, this relation may
be determined either by the mechanism responsible for the nearly
complete loss of angular momentum required by the gas to accrete onto
the BH, or by a self regulation of the bulge-BH system.  To clarify
this point we have shown results based on the novel galaxy formation
model of Monaco et al. (2006), where a model with no winds is unable
to fit the hard-X LF of AGNs, while models with dry or accreting winds
are more successfull.  However, all the models reproduce the BH--bulge
relation in a similar way.

Finally, the most promising mechanisms to achieve the downsizing of
AGNs are found to be from the one hand the quenching of late cooling
flows by AGN jets, able to avoid the late accretion of mass by massive
bulges/BHs, on the other hand the stellar kinetic feedback that takes
place in star-forming bulges, able to delay the assembly of small
bulges at high redshift.

\begin{acknowledgements}
I thanks my collaborators Fabio Fontanot, Stefano Cristiani, Laura
Silva, Eros Vanzella, Mario Nonino and Paolo Tozzi for their help.

\end{acknowledgements}

\bibliographystyle{aa}

\begin{thebibliography}{}


\bibitem[Begelman(2004)]{2004cbhg.symp..375B} Begelman M.C., 2004, in L.C. Ho ed., Coevolution of Black Holes and Galaxies. Cambridge University Press, Cambridge, p. 375
\bibitem[]{ac} Begelman M.C., McKee C.F., Shields G.A., 1983, ApJ, 271, 70
\bibitem[]{bo} Bower, R. G., Benson, A. J., Malbon, R., Helly, J. C., Frenk, C. S., Baugh, C. M., Cole, S., Lacey, C. G., 2005, astro-ph/0511338
\bibitem[]{ad} Bromley J.M., Somerville R.S., Fabian A.C., 2004, MNRAS, 350, 456
\bibitem[]{af} Cavaliere A., Vittorini V., 2002, ApJ, 570, 114 
\bibitem[Ciotti \& Ostriker(1997)]{1997ApJ...487L.105C} Ciotti L., Ostriker J.P., 1997, ApJ, 487, L105
\bibitem[]{ah2} Cristiani S., et al., 2004, ApJL, 600, 119
\bibitem[]{ao} Fabian A.C., 1999, MNRAS, 308, 39
\bibitem[]{ff} Fontanot, F., Cristiani, S., Monaco, P. et al., 2006a, in preparation
\bibitem[]{fa} Fontanot, F., Monaco, P., Cristiani, S. et al., 2006b, in preparation
\bibitem[]{ap} Granato G.L., De Zotti G., Silva L., Bressan A., Danese L., 2004, ApJ, 600, 580
\bibitem[]{aq} Granato G.L., Silva L., Monaco P., Panuzzo P., Salucci P., De Zotti G., Danese L., 2001, MNRAS, 324, 757
\bibitem[]{ar} Haehnelt M.G., Natarajan P., Rees M.J., 1998, MNRAS, 300, 817
\bibitem[]{ha} Hasinger, G., Miyaji, T., \& Schmidt, M.\ 2005, \aap, 441, 417
\bibitem[]{av} Kauffmann G., Haehnelt M.G., 2000, MNRAS, 311, 576
\bibitem[]{king} King, A.\ 2005, \apj, in press (astro-ph/0511034)
\bibitem[La Franca et al.(2005)]{2005ApJ...635..864L} La Franca, F., et al.\ 2005, \apj, 635, 864 
\bibitem[]{lcm} Lapi, A., Cavaliere, A., Menci, N.\ 2005, \apj, 619, 60
\bibitem[]{bj} Mahmood A., Devriend J.E.G., Silk J., astro-ph/0401003
\bibitem[]{cbx}Marconi, A., Risaliti, G., Gilli, R., Hunt, L.~K., Maiolino, R., Salvati, M. 2004, MNRAS, 351, 169
\bibitem[]{me2} Merloni, A., 2004, MNRAS, 353, 1035
\bibitem[]{bs} Monaco P., 2004, MNRAS, 352, 181
\bibitem[]{mfo} Monaco, P., Fontanot, F. 2005, \mnras, 359, 283
\bibitem[]{gal} Monaco, P., Fontanot, F., Taffoni, G., 2006, in preparation
\bibitem[Murray et al.(2005)]{2005ApJ...618..569M} Murray N., Quataert E., Thompson T.A., 2005, ApJ, 618, 569
\bibitem[]{bz} Ostriker J.P., McKee C.F., 1988, RvMP,60, 1O
\bibitem[]{yu} Sazonov S.Yu., Ostriker J.P., Ciotti L., Sunyaev R.A., 2005, MNRAS, 358, 168
\bibitem[]{ch} Silk J., Rees M.J., 1998, A\&A, 331, 1
\bibitem[]{treu} Treu, T., Ellis, R. S., Liao, T. X., van Dokkum, P. G. 2005, \apj, 622, L5
\bibitem[]{hxo}Ueda Y., Akiyama M., Ohta K., Miyaji T., 2003, ApJ, 598, 886

\end{thebibliography}

\end{document}